\documentclass[journal]{IEEEtran}
\ifCLASSINFOpdf
\else
\fi
\usepackage{bm}
\usepackage{amsmath}
\usepackage{booktabs}
\usepackage{graphicx}
\usepackage{subfigure}
\usepackage{makecell}
\usepackage{array}
\usepackage{color}
\usepackage{verbatim}
\usepackage{chngcntr}
\usepackage{float}
\begin{document}
\title{Forward and inverse problems for Eikonal equation based on DeepONet}
%
%
%

\author{Yifan Mei,
        Yijie Zhang,
        Xueyu Zhu,
        and Rongxi Gou
\thanks{Y. Mei, Y. Zhang and R. Gou are with the School of Information and Communications Engineering, Xi'an Jiaotong University, Xi'an, Shaanxi 710049, China (e-mail: meiyf1028@163.com; zhangyijie2016@mail.xjtu.edu.cn; grx123456@stu.xjtu.edu.cn).}
\thanks{X. Zhu is with the Department of Mathematics, The University of Iowa, Iowa City, IA 52246 (e-mail: xueyu-zhu@uiowa.edu).}}

%
%

\markboth{IEEE GEOSCIENCE AND REMOTE SENSING LETTERS,~2023}%
{Shell \MakeLowercase{\textit{et al.}}: Bare Demo of IEEEtran.cls for IEEE Journals}
%



\maketitle

\begin{abstract}
Seismic forward and inverse problems are significant research areas in geophysics.
However, the time burden of traditional numerical methods hinders their applications in scenarios that require fast predictions. 
Machine learning-based methods also have limitations as retraining is required for every change in initial conditions.
In this letter, we adopt deep operator network (DeepONet) to solve forward and inverse problems based on the Eikonal equation, respectively.
DeepONet approximates the operator through two sub-networks, branch net and trunk net, which offers good generalization and flexibility.
Different structures of DeepONets are proposed to respectively learn the operators in forward and inverse problems.
We train the networks on different categories of datasets separately, so that they can deliver  accurate predictions with different initial conditions for the specific velocity model.
The numerical results demonstrate that DeepONet can not only predict the travel time fields with different sources for different velocity models, but also provide  velocity models based on the observed travel time data.
\end{abstract}


%
\IEEEpeerreviewmaketitle

\section{Introduction}
%
%
%
%
\IEEEPARstart{S}{eismic} forward modeling and inversion are important methods for exploring the structure of the underground.
Fermat’s principle states that the path taken by a ray between two given points is the one that minimizes the travel time.
A well-known partial differential equation (PDE) in seismic travel time problems is the eikonal equation, which is the high-frequency approximation of the wave equation.
There are several numerical methods for solving the eikonal equation including fast marching method \cite{sethian19993}, finite difference method \cite{qin1992finite} and fast sweeping method \cite{zhao2005fast}.
But the large amount of computation required to numerically solve the PDE for complex systems hinders its application in practical problems that require fast prediction.
In practice, system parameters are often difficult to measure directly and need to be calculated through indirect measurements. Traditional methods for solving inverse problems often require repeatedly solving the forward problem, which means it can be time-consuming.
\par
With the development of machine learning (ML) and computational capacity, there is a growing interest in utilizing ML to solve forward and inverse problems based on PDEs \cite{moseley2018fast}, \cite{adler2021deep}, in order to overcome the shortcomings of traditional methods.
In the past few years, many research communities have concentrated their efforts on neural operators \cite{li2020fourier,lu2019deeponet, 2021Neural, patel2021physics}, which can learn  the physical systems at the operator level.
Notably, deep operator network (DeepONet) \cite{lu2019deeponet}  has become one of the preferred methods for solving PDEs due to its fast learning speed, low generalization error, and flexibility of the structure.
DeepONet approximates the operator by representing the base of the mapping and the coefficients of the base through two sub-networks (branch net and trunk net).
And its two sub-networks are not limited to any specific architecture and can be implemented using FNN, CNN or other types of networks.
In the network training phase, DeepONet obtains the appropriate mapping in the data by learning the parameters of the two sub-networks.
Recently, DeepONet's ability to quickly and accurately simulate complex dynamics has attracted attention in many fields, including fluid mechanics \cite{lin2021seamless},  \cite{lin2021operator}, hypersonics \cite{mao2021deepm}, \cite{zanardi2022towards}, combustion \cite{ranade2021generalized} and biomechanics \cite{yin2022simulating}.
In order to approximate multiple outputs, impose boundary conditions, and improve the accuracy of DeepONet, some new extensions are proposed \cite{lu2022comprehensive}.

In this letter, we adopt DeepONet to solve forward and inverse problems for eikonal equations. 
Specifically, we propose several DeepONets with different structures for different problem settings.
We train network models on several different categories of datasets, and then input data (such as travel time or velocity) to the network for rapid prediction.
For the forward problem, the neural networks can quickly and accurately predict the travel time of different sources on different velocity models.
For the inverse problem, the neural networks can provide a reasonably accurate velocity field based on observed travel time data without any prior knowledge about the velocity field. 
Particularly, instead of necessarily retraining the network as long as the inputs are changed \cite{bin2021pinneik}, our method can predict results for the specific category of the velocity model once the network is trained.
\par
This letter is organized as follows. In Section \ref{sec:method}, we briefly introduce the eikonal equation. Then we describe the forward problem and the inverse problem and the implementation of the basic structures of DeepONet for both of these problems. In Section \ref{sec:result}, we present several numerical examples to demonstrate the effectiveness of the method. Finally, we conclude in Section \ref{sec:conclusion}.

\section{Method}
\label{sec:method}
\subsection{Eikonal equation}
The eikonal equation is the basic equation in geophysics to describe the relationship between the first arrival time and the velocity of wave propagation. 
The eikonal equation in two-dimensional variable-velocity media can be expressed as follow:
\begin{equation}
\label{eq:eikonal}
\left\{\begin{array}{l}
\left|\nabla T\left(\bm{x}_s, \bm{x}\right)\right|^2=\displaystyle\frac{1}{v^2(\bm{x})}, \forall \bm{x} \in \Omega \\
T\left(\bm{x}_s, \bm{x}_s\right)=0
\end{array}\right.
\end{equation}
where $T(\bm{x}_s,\bm{x})$ represents the travel time from the source point $\bm{x}_s$ to any point $\bm{x}$ in $\Omega$, $v(\bm{x})$ represents the velocity defined in $\Omega$.
$T(\bm{x}_s,\bm{x})$ can be factorized into two factors as follows:
\begin{equation}
\label{eq:two_factors}
\left\{\begin{array}{l}
T\left(\bm{x}_s, \bm{x}\right)=T_0\left(\bm{x}_s, \bm{x}\right) \tau\left(\bm{x}_s, \bm{x}\right), \\
T_0\left(\bm{x}_s, \bm{x}\right)=\displaystyle\frac{\left|\bm{x}-\bm{x}_s\right|}{v\left(\bm{x}_s\right)},
\end{array}\right.
\end{equation}

The goal of the forward problem is to predict travel-time $T(\bm{x}_s, \bm{x})$ from the specific model $v(\bm{x})$. For inverse problem, we aim to predict the velocity fields $v(\bm{x})$ based on the observed travel-time $T(\bm{x}_s, \bm{x})$ from a set of specific locations.

\subsection{Forward and inverse problems for eikonal equation}
Deep Operator Network (DeepONet) was proposed by Lu et al. \cite{lu2021learning}, which learns the operator from the data based on two sub-networks: the branch net for the input function and the
trunk net for the locations to evaluate the output function.
\subsubsection{Forward problem}
In the forward problem, our goal is to approximate the travel time from the velocity field, in other words, the operator that DeepONet needs to learn is $\mathcal{G}_{for}:v(\bm{x}) \mapsto T\left(\bm{x}_s, \bm{x}\right)$. 
From equation \eqref{eq:two_factors}, once  $v$ and $\Omega$ are given, $T_0\left(\bm{x}_s, \bm{x}\right)$ can be calculated immediately, so the network actually needs to learn the operator $\mathcal{G}_{for}:v(\bm{x}) \mapsto \tau\left(\bm{x}_s, \bm{x}\right)$.
\par 
We utilize the velocity models $v(\bm{x})$ as the input of the branch net, and also add $\bm{x}_s$ into the branch net as additional information.
The trunk net has only one input $\bm{x}$ that represents the locations of the output of DeepONet.
So the DeepONet for the forward problem can be represented as
\begin{equation}
\begin{split}
&\mathcal{N}_{\rm{for}}\left(v(\bm{x}), \bm{x}_s, \bm{x} ; \theta_b^{\rm{for}},\theta_t^{\rm{for}}\right) \\ 
&= \underbrace {\mathcal{N}_b\left(v(\bm{x}), \bm{x}_s ; \theta_b^{\rm{for}}\right)}_{\rm{branch}}
\underbrace {\mathcal{N}_t\left(\bm{x} ; \theta_t^{\rm{for}}\right)}_{\rm{trunk}} + b^{\rm{for}} ,
\end{split}
\label{eq:forward}
\end{equation}
where $\theta_b^{\rm{for}}$ and $\theta_t^{\rm{for}}$ represent the weights and bias of branch net and trunk net, respectively, $b^{\rm{for}}$ is a bias added in the last stage.
\par
\begin{figure}
    \centering    \includegraphics[width=1\linewidth]{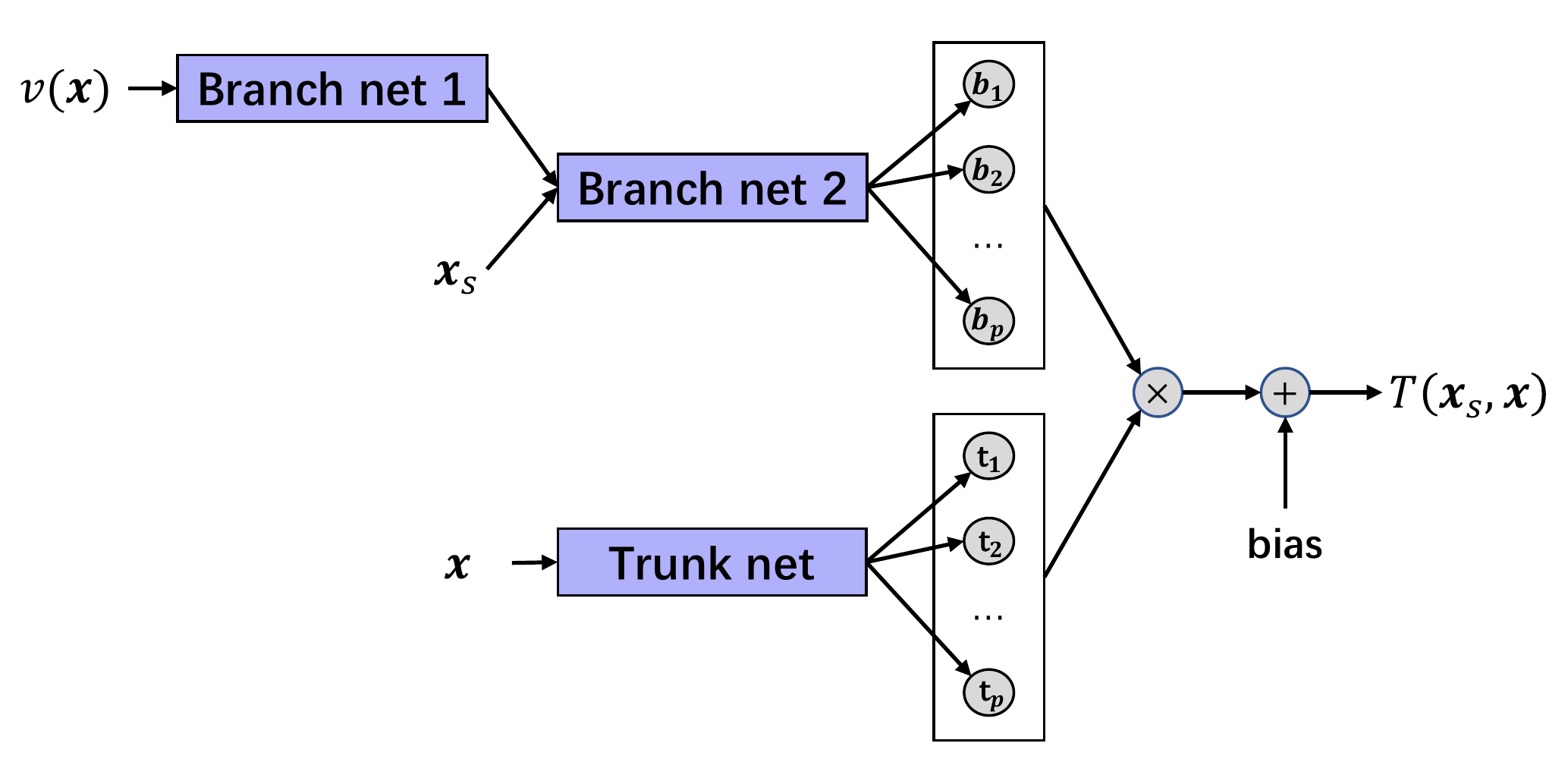}
    \caption{The structure of DeepONet for the forward problem. The functions $v(\bm{x})$ are fed into branch net 1, and the source coordinates $\bm{x}_s$ are combined with the output of branch net 1 and fed into branch net 2. The coordinates of the predicted point $\bm{x}$ are input into the trunk net.}
    \label{fig:forward}
\end{figure}

We divide the branch net into two sub-networks to better accommodate the source locations $\bm{x}_s$. 
Specifically, velocity models are fed into branch net 1, while source locations are paired with the output of branch net 1 and fed into branch net 2. This design not only enables the model to effectively capture the correlation between velocity models and source locations, but also prevents the larger size of velocity models from dominating the learning process and overshadowing the importance of source locations. 
Moreover, the output of branch net 1 can be viewed as a feature extraction process for velocity models, thus facilitating a better understanding of the inter-relationship between velocity models and source locations by branch net 2.
The basic structure of the neural network is shown in Figure \ref{fig:forward}.

\subsubsection{Inverse problem}
For the inverse problem, our goal is to approximate the velocity field based on the observed travel time. 
The operator that DeepONet needs to learn is $\mathcal{G}_{inv}: T\left(\bm{x}_s, \bm{x}\right) \mapsto v(\bm{x})$. 

\par
The input functions for the branch net in our approach are the observed travel time, while the trunk net only has one input $\bm{x}$.
The DeepONet can be represented as follows:
\begin{equation}
\begin{split}
&\mathcal{N}_{\rm{inv}}\left(T(\bm{x}_s,\bm{x}), \bm{x} ; \theta_b^{\rm{inv}},\theta_t^{\rm{inv}}\right)\\ 
&= \underbrace {\mathcal{N}_b\left(T(\bm{x}_s,\bm{x}) ; \theta_b^{\rm{inv}}\right)}_{\rm{branch}}
\underbrace {\mathcal{N}_t\left(\bm{x} ; \theta_t^{\rm{inv}}\right)}_{\rm{trunk}} + b^{\rm{inv}} ,
\end{split}
\end{equation}
where $T(\bm{x}_s,\bm{x})$ represents the observed travel time. 
The other parameters are similar to \eqref{eq:forward}.
\par
\begin{figure}
    \centering    \includegraphics[width=0.8\linewidth]{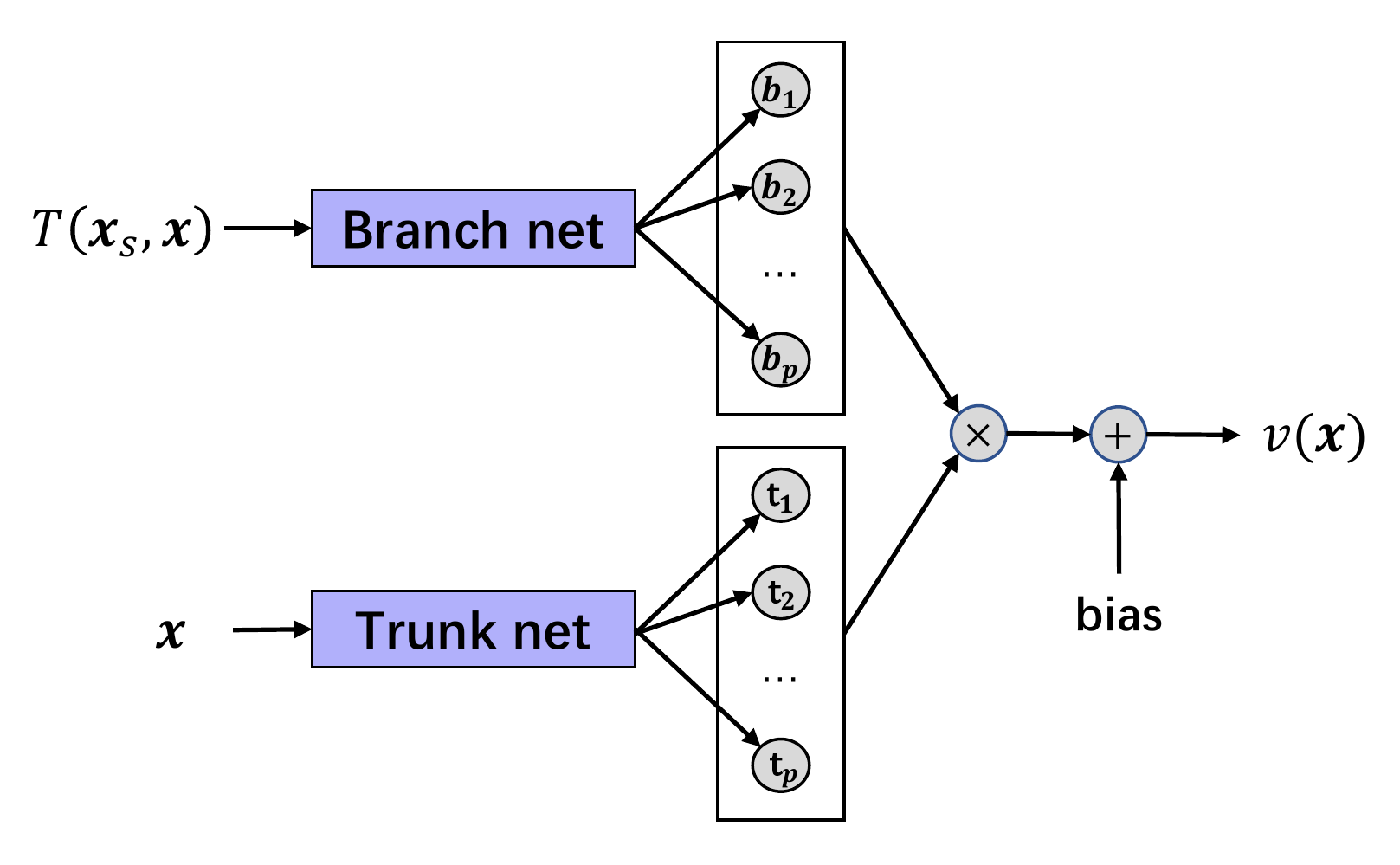}
    \caption{The structure of DeepONet for inverse problem. The functions $T(\bm{x}_s,\bm{x})$ are fed into branch net and the coordinates of the predicted point $\bm{x}$ are input into the trunk net.}
    \label{fig:inverse}
\end{figure}

The basic structure of the neural network is shown in Figure \ref{fig:inverse}. 

\subsubsection{Loss function}
Both in forward and inverse problems, we utilize the mean square error (MSE) as the loss function, which can be expressed as:
\begin{equation}
\mathcal{L}_{\rm{for}}(\theta_b^{\rm{for}},\theta_t^{\rm{for}}; T, \hat{T})=\frac{1}{N} \sum_{i=1}^N\left(T_i-\hat{T}_i\right)^2
\end{equation}
\begin{equation}
\mathcal{L}_{\rm{inv}}(\theta_b^{\rm{inv}},\theta_t^{\rm{inv}}; v, \hat{v})=\frac{1}{N} \sum_{i=1}^N\left(v_i-\hat{v}_i\right)^2
\end{equation}
where $N$ is the batch size, $i$ represents the 
$i$-th sample. ${\hat{T}}$, $\hat{v}$ are the approximate solutions for the forward and inverse problems, respectively. $T$ and $v$ are the ground truth values corresponding to ${\hat{T}}$ and $\hat{v}$. 

\section{Result}
\label{sec:result}
In this section, we utilize 10 types of velocity models from the OpenFWI dataset \cite{deng2021openfwi}  to demonstrate the performance of DeepONet for forward problems, which includes FlatVel-A, FlatVel-B, CurveVel-A, CurveVel-B, FlatFault-A, FlatFault-B, CurveFault-A, CurveFault-B, Style-A, Style-B.
In addition, we choose FlatVel-A and Style-A to demonstrate the performance of DeepONet on inverse problems.
For the above datasets, the spatial domain is defined on a 70 × 70 grid with $\Delta x$ = $\Delta z$ = 20 m.
In order to facilitate the training of  the neural network, we utilize km/s and km as the units of the velocity and the model size, respectively.
For the  numerical examples, we utilize a toolkit called scikit-fmm \cite{furtney2021scikit} to generate travel time data 
from velocity models.
\par
To quantify the performance of our method, we use the relative error (RE) and the standard deviation (STD) of RE as
\begin{equation}{\rm{RE}}=\frac{1}{n} \sum_{i=1}^{n}{\rm{RE}}_i,\end{equation}
\begin{equation}{\rm{STD}}=\sqrt{\frac{1}{n} \sum_{i=1}^n\left({\rm{RE}}_i-\rm{RE}\right)^2},\end{equation}
here,
\begin{equation}{\rm{RE}}_i=\sqrt{\frac{\sum_{j=1}^m\left|y_j^i-\hat{y}_j^i\right|^2}{\sum_{j=1}^m\left|y_j^i\right|^2}},\end{equation}
where $n$ represents the number of test examples in the test set, and $m$ represents the number of test points in a single test example.
$\hat{y}$ is the output results of DeepONet, $y$ is the ground truth corresponding to $\hat{y}$.
${\rm{RE}}_i$ is the relative error of the $i$-th test example.

Our neural networks are trained by the Pytorch on an NVDIA GeForce RTX 3090 GPU.


\subsection{Forward problem}
For the forward problem based on the eikonal equation, we train the network on 10 types of velocity models and test the performance of the network for each of them separately. 
In addition, we randomly select 300 velocity models as the training set and 100 velocity models as the test set for each category \cite{deng2021openfwi}.
The DeepONet for the forward problem is trained using the Adam optimizer with a learning rate of 1e-5, and the learning rate is decayed by half when the training loss does not decrease for 10 epochs. 
The training process consists of 300 epochs, and each epoch includes a batch size of 980 samples.

\par
For each velocity model, two source locations are randomly selected to allow the network to learn more information.
For both the training and test sets, the sources are located at the  surface ($z_s = 0$ km).
Therefore, for the  source locations $\bm{x}_s=(x_s,z_s)$, we set $z_s = 0$ and randomly choose two values for $x_s$.
To reduce the amount of data that needs to be processed during each training iteration, we discretize the velocity model along the $x$-axis by increasing $\Delta x$ = 20 m to $\Delta x$ = 100 m. 
The details of the neural network are shown in Table \ref{tab:forward_layers}.
\par
The results of the experiments are shown in Table \ref{tab:forward}, which presents the RE and STD of the network on the test set. For the velocity models with relatively simple structures, such as FlatVel-A, CurveVel-A, FlatFault-A, and CurveFault-A, the neural network has significantly smaller errors on the test set.

\begin{table}[htb]
    \centering
    \caption{The relative error and standard deviation of DeepONet on the test set for the forward problem.}
    \begin{tabular}{ccc}
    \toprule
     & RE & STD \\
    \midrule 
    FlatVel-A &0.013&0.005  \\
    FlatVel-B &0.045&0.021  \\
    CurveVel-A &0.042&0.020  \\
    CurveVel-B &0.072&0.044   \\
    FlatFault-A &0.022&0.011  \\
    FlatFault-B &0.094&0.039  \\
    CurveFault-A &0.040&0.028  \\
    CurveFault-B &0.146&0.078  \\
    Style-A &0.072&0.042  \\
    Style-B &0.071&0.037  \\
    \bottomrule
    \end{tabular}
    \label{tab:forward}
\end{table}

\begin{figure}[htb]
    \centering
    \subfigure[Stacked histogram of the error distribution for velocity models of A Family.]{
    \includegraphics[width=0.60\linewidth]{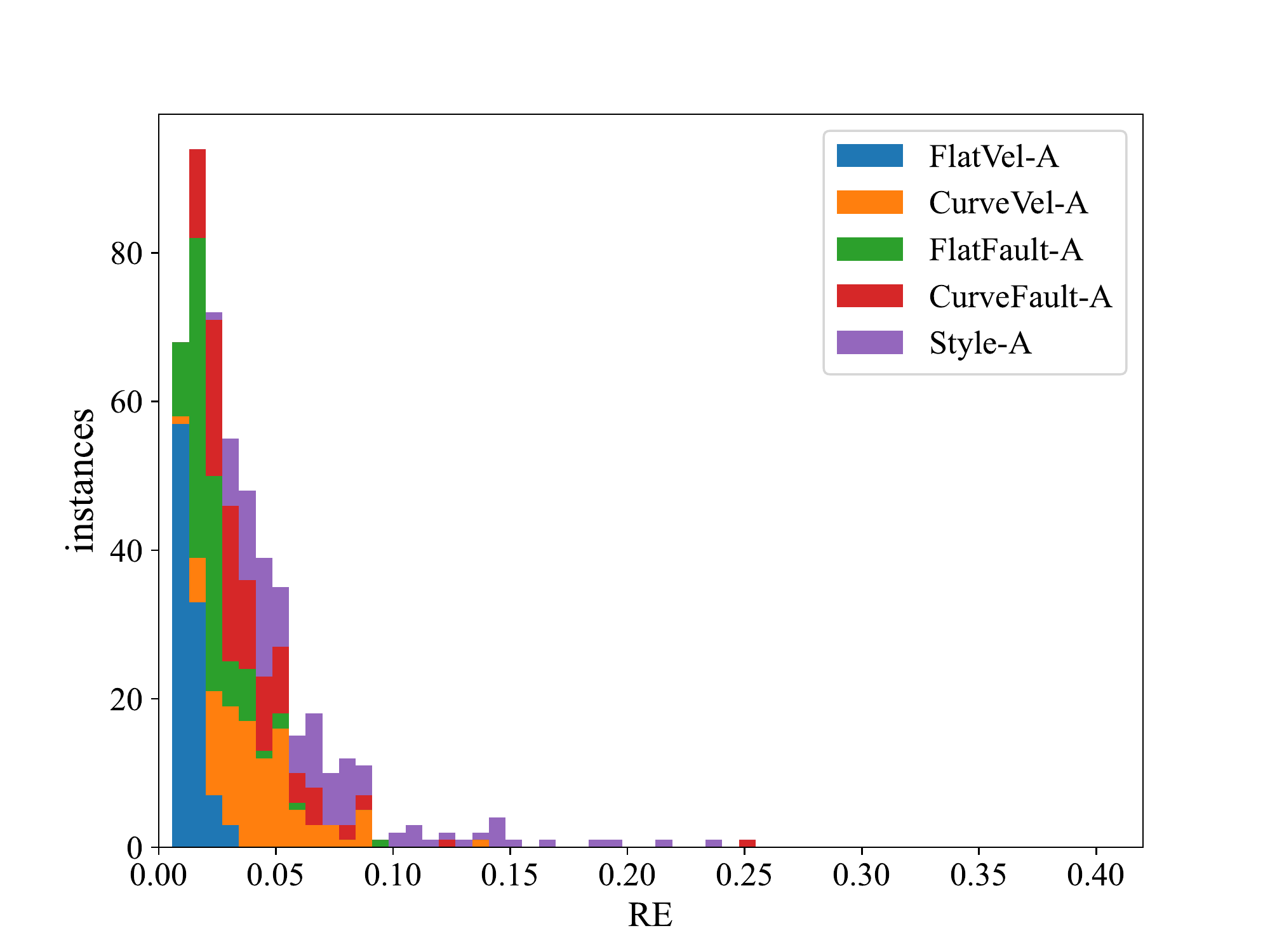}
    }
    \subfigure[Stacked histogram of the error distribution for velocity models of B Family.]{
    \includegraphics[width=0.60\linewidth]{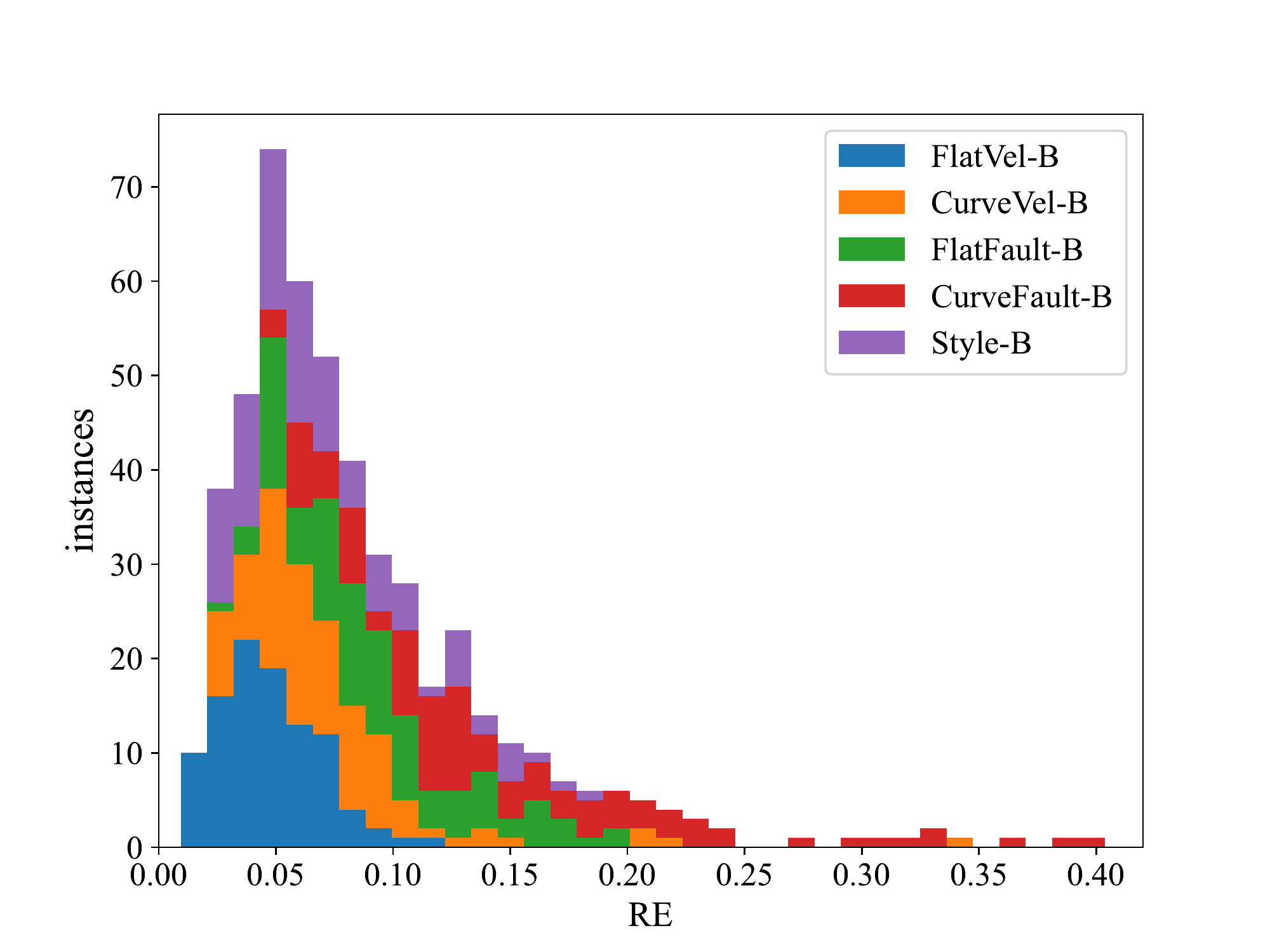}
    }
    \caption{
    Stacked histogram of the error distribution in the test set for the forward problem by the DeepOnet.
    The horizontal axis represents error range, the vertical axis is frequency of corresponding errors in the range.
    }
    \label{fig:histogram_error_branch}
\end{figure}

Figure \ref{fig:histogram_error_branch} shows the stacked histogram of the error distribution in the test set, where the relative error of 95.6\% instances in Family A is less than 0.1 and the relative error of 70.8\% instances in Family B is less than 0.1 based on the DeepOnet as in Figure \ref{fig:forward}, which demonstrates the accuracy of the proposed method.
In addition, the average prediction time for predicting travel-time is 0.84$\pm$0.04 ms, 
which illustrates the efficiency of DeepONet for the forward problem. 

\subsection{Inverse problem}
Next, we test the performance of the DeepONet for the inverse problem on velocity models: FlatVel-A and Style-A.
When training on each category of velocity models separately, we choose 1000 velocity models from each category as the training set and 300 velocity models as the test set.

\begin{figure}[htb]
    \centering
    \subfigure[The ground truth model. (FlatVel-A)]{
    \includegraphics[width=0.46\linewidth]{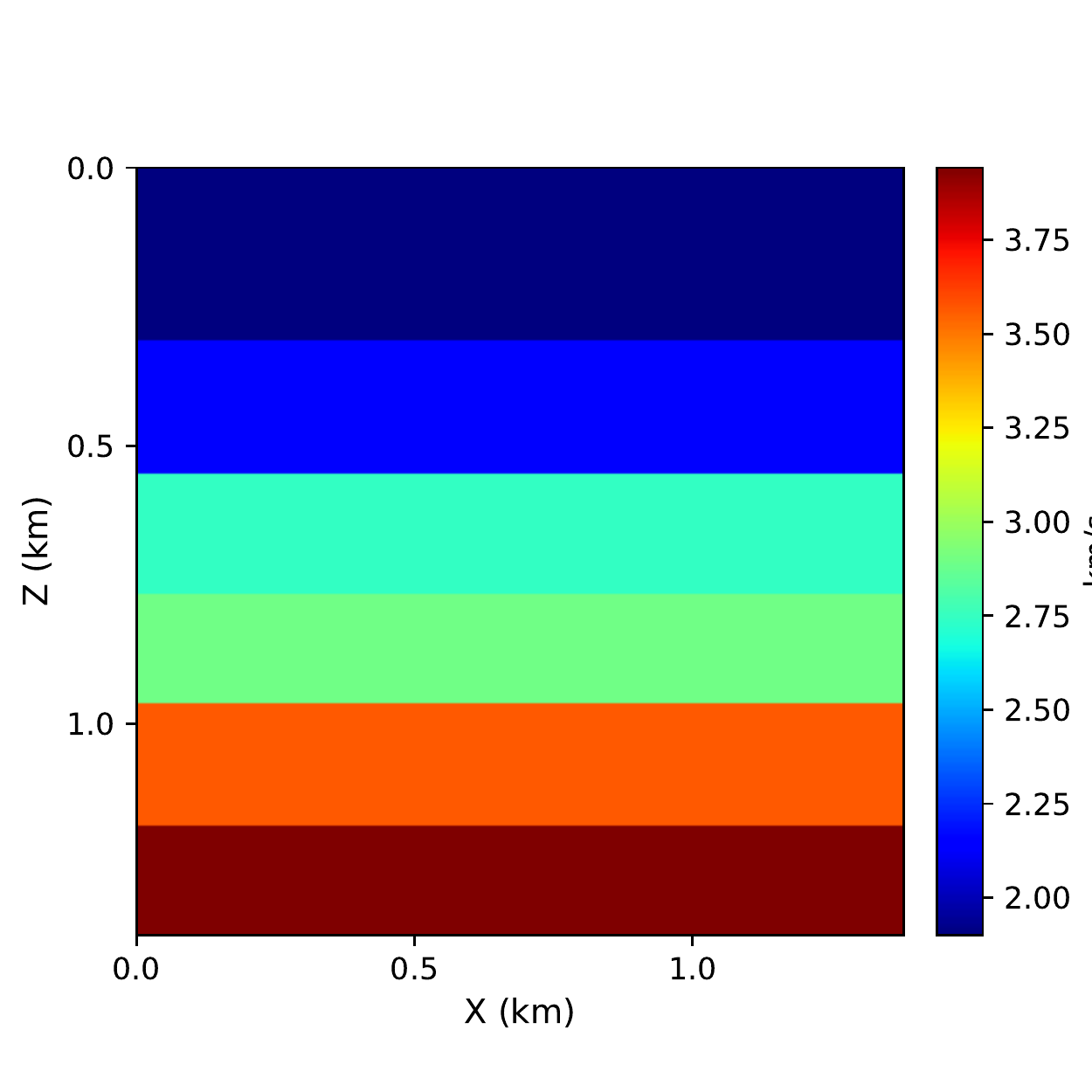}
    }
    \subfigure[The predicted model. (FlatVel-A)]{
    \includegraphics[width=0.46\linewidth]{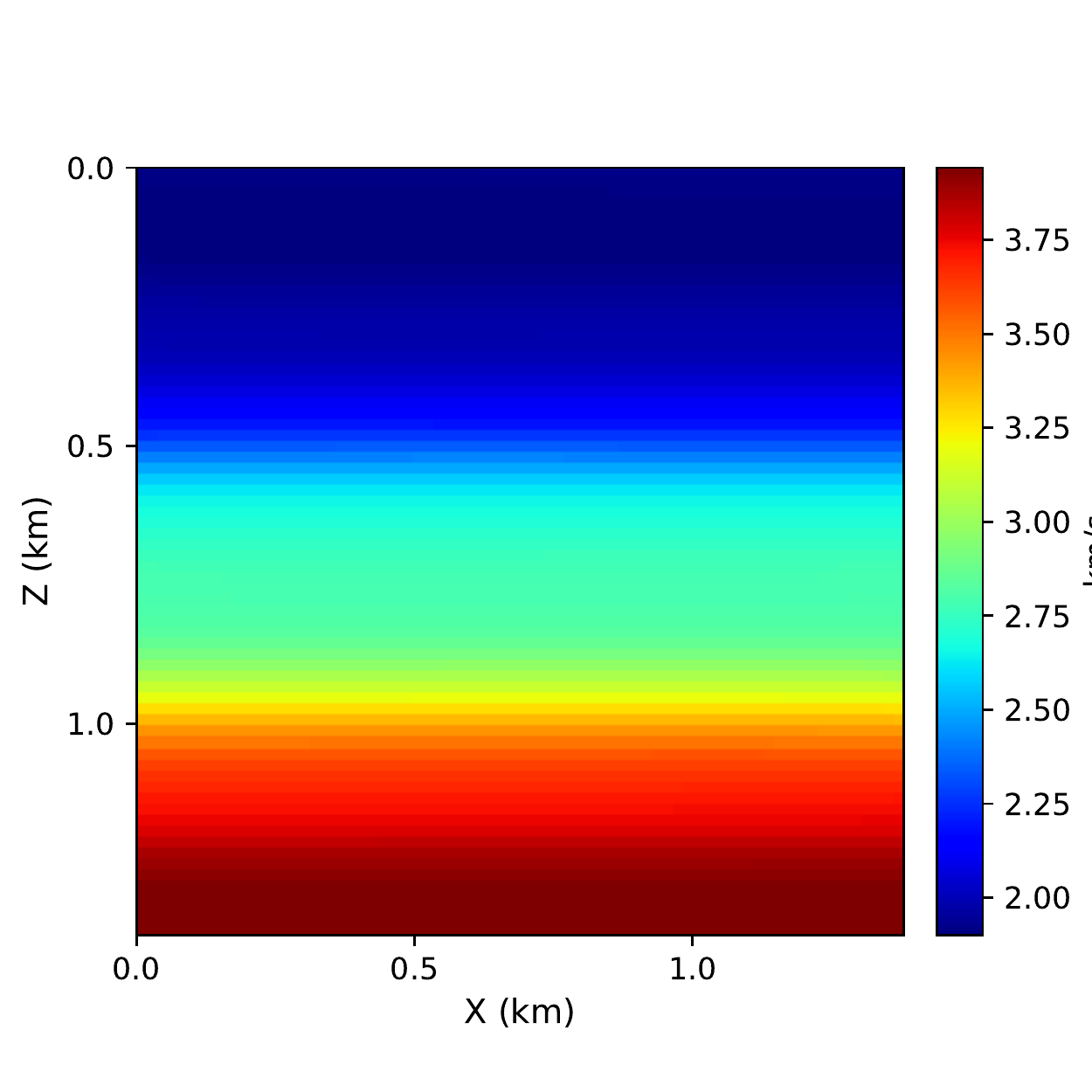}
    }
    \subfigure[The ground truth model.(Style-A)]{
    \includegraphics[width=0.46\linewidth]{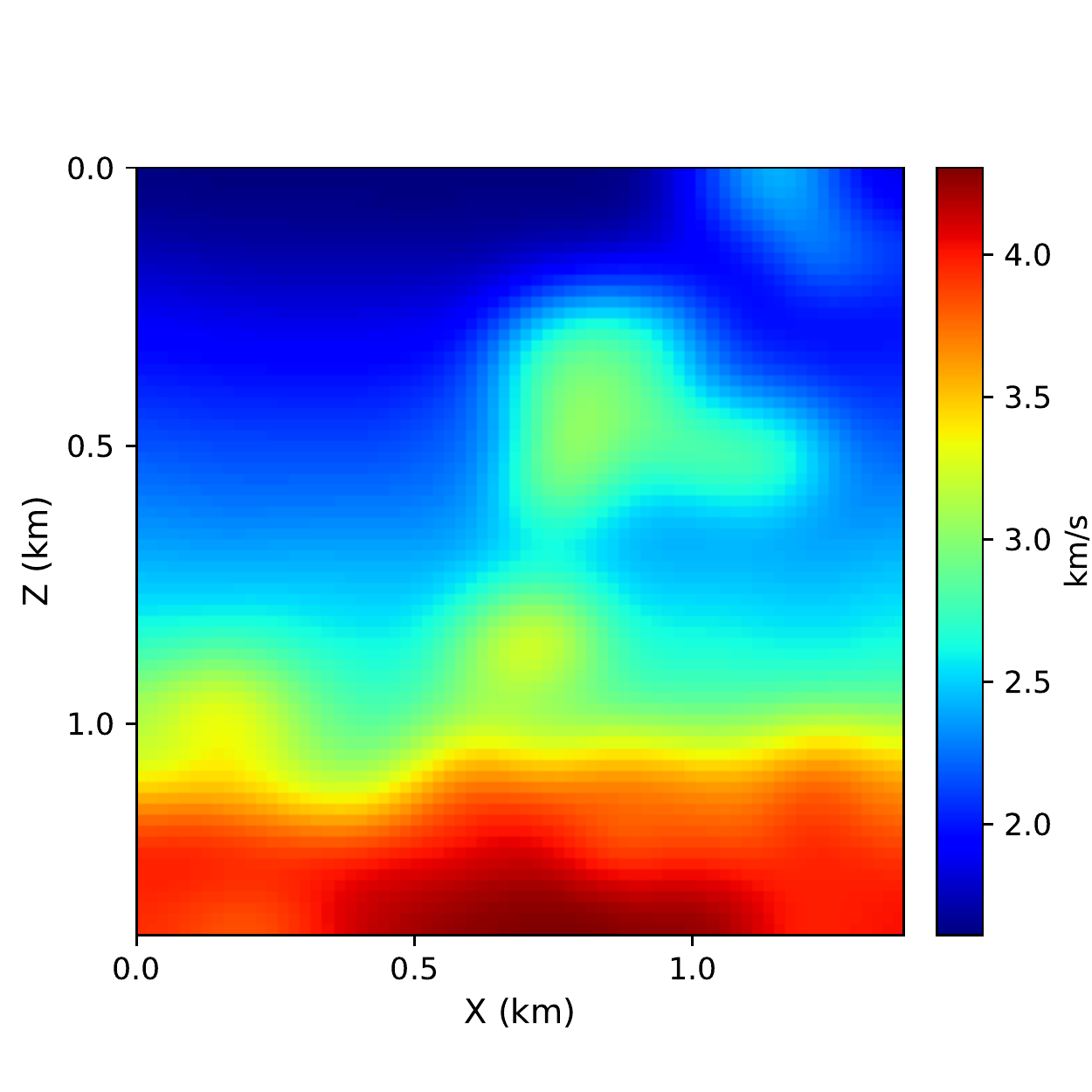}
    }
    \subfigure[The predicted model.(Style-A)]{
    \includegraphics[width=0.46\linewidth]{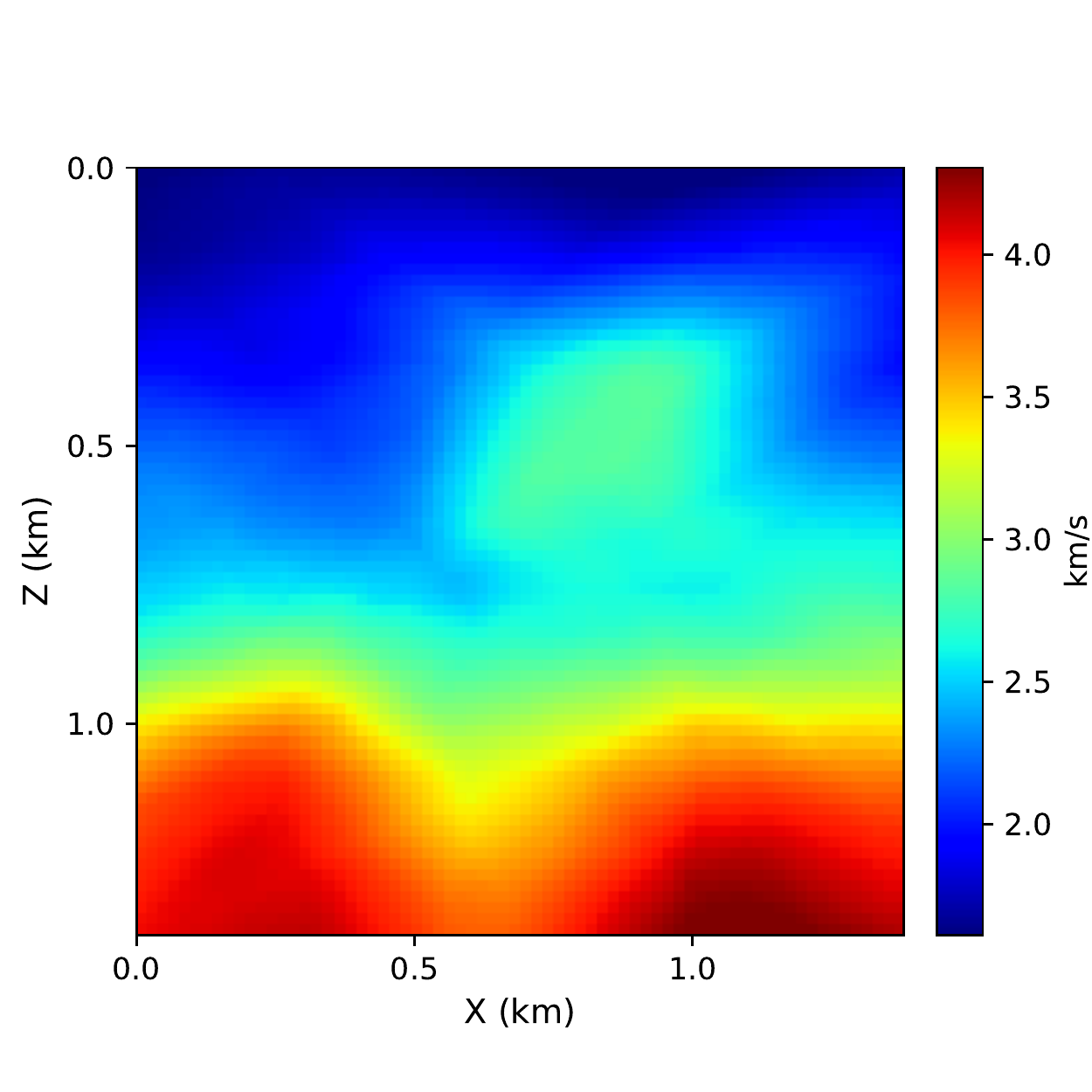}
    }
    \caption{
    (a) The ground truth model in FlatVel-A, and (b) the predicted velocity model. (c) The ground truth model in Style-A, and (d) the corresponding predicted velocity model.
    }
    \label{fig:result_inverse}
\end{figure}

As shown in Figure \ref{fig:result_inverse} (a) and (c), considering the different levels of structural complexity, we adopt different  numbers of  sources for the FlatVel-A and Style-A models.
The FlatVel-A velocity model in Figure \ref{fig:result_inverse} (a), which has a layered structure, and the velocity field only varies in depth. The structure is relatively simple, but there is a discontinuous change in velocity between layers.
We place 7 equally spaced sources along the left boundary ($x=$ 0 km), and 70 receivers along the right boundary ($x=$ 1.38 km) to record arrival time data.
For the Style-A velocity model in Figure \ref{fig:result_inverse} (b), which is rather complex. Although the velocity field varies smoothly, there are irregular variations in both the vertical and horizontal directions. We increase the number of sources on the left boundary to 10, while keeping the configuration of receivers on the right boundary. 
we still discretize the velocity model along the $x$-axis with $\Delta x$ = 100 m instead of 20 m. 

\par

The DeepONet for the inverse problem is trained by Adam optimizer with 300 epochs on FlatVel-A model and 400 epochs on Style-A model. 
The network is trained using a learning rate of 5e-4.
If the training loss fails to decrease for 20 epochs, the learning rate is reduced by a factor of 0.5.
The other hyperparameters are the same as DeepONet for the forward problem.
\par

The details of each layer of DeepONet for the inverse problem of FlatVel-A and Style-A are shown in Table \ref{tab:inverse_layers_flatvel_a} and Table \ref{tab:inverse_layers_style_a}, respectively.
\par
Figure \ref{fig:result_inverse} shows the predicted velocity models on the test set. 
Figure \ref{fig:result_inverse} (a) is a model in FlatVel-A, which is a multi-layer structure. 
Figure \ref{fig:result_inverse} (b) is the predicted velocity model. It can be seen that the main structure of the velocity field are basically consistent with Figure \ref{fig:result_inverse} (a), except for the locations with significant jumps. 
Figure \ref{fig:result_inverse} (c) is a model in Style-A, which has a smooth but complex structure. Figure \ref{fig:result_inverse} (d) is the corresponding predicted model by DeepONet. 
We observe that the network can capture the main characteristics of the velocity field, but significant discrepancies exist in the details compared to the ground truth model.
The RE and STD of the trained neural network on the test set are shown in Table \ref{tab:inverse}.

\begin{table}[htb]
    \centering
    \caption{The relative error and standard deviation of DeepONet on the test set for the inverse problem.}
    \begin{tabular}{ccc}
    \toprule  
     & RE & STD \\
    \midrule  
    FlatVel-A & 0.053 & 0.016  \\
    Style-A & 0.102 & 0.027  \\
    \bottomrule  
    \end{tabular}
    \label{tab:inverse}
\end{table}

The average prediction time for velocity model prediction in FlatVel-A and Style-A is 1.08$\pm$0.02 ms and 1.77$\pm$0.03 ms, respectively, which 
illustrates the efficiency of DeepONet for inverse problems.

\par

\section{Conclusion}
\label{sec:conclusion}
In order to mitigate the computation cost of conventional numerical methods in solving both forward and inverse problems, we introduced DeepONet based on the eikonal equation.
Due to the differences in the forward and inverse problems, especially the input data, we proposed two structures of DeepONet to learn the corresponding operators, respectively.
In the forward problem, we tested the performance of DeepONet on 10 categories of velocity models separately. 
For the inverse problem, we tested the performance of the network with limited observed travel-time data as input on two types of velocity models, respectively. 
Several numerical experiments showed that DeepONet predicts travel time quickly and accurately for different sources on different velocity models. In addition, DeepONet provides a reasonably accurate velocity field based on the observed travel time data.


%

\appendices
\counterwithin{table}{section}
\section{Details of the layers in DeepONet}
In appendix, we list the structures of the DeepONet for forward and inverse problems, respectively.
\begin{table*}[!htbp]
    \renewcommand{\arraystretch}{1}
    \centering
    \caption{Details of the layers in DeepONet for the forward problem.}
    \label{tab:forward_layers}
    \scriptsize
    \begin{tabular}{ccc|ccc|ccc}
    \hline
    \multicolumn{3}{c|}{branch net 1} & \multicolumn{3}{c|}{branch net 2} & \multicolumn{3}{c}{trunk net} \\
    \hline
    layer & in\_features & out\_features & layer & in\_features & out\_features & layer & in\_features & out\_features \\
    \hline
    linear+relu  & 4900  & 2500  & linear+relu    & 5     & 50 & linear+relu & 2 & 50 \\
    linear+relu  & 2500  & 500   & linear+relu    & 50    & 50 & linear+relu & 50 & 50 \\
    linear+relu  & 500   & 100   & linear+relu    & 50    & 5  & linear+relu & 50 & 5 \\
    linear+relu  & 100   & 50    &      &  &  &  &  &  \\
    linear+relu  & 50    & 5     &      &  &  &  &  &  \\
    linear+relu  & 5     & 3     &      &  &  &  &  &  \\
    \hline 
    \end{tabular}
\end{table*}

\begin{table*}[!htbp]
    \renewcommand{\arraystretch}{1}
    \centering
    \caption{Details of the layers in DeepONet for the inverse problem of FlatVel-A.}
    \label{tab:inverse_layers_flatvel_a}
    \scriptsize
    \begin{tabular}{ccccc|ccc}
    \hline
    \multicolumn{5}{c|}{branch net} & \multicolumn{3}{c}{trunk net} \\
    \hline
    layer & \makecell[c]{in\_channels\\/in\_features} & \makecell[c]{out\_channels\\/out\_features} & kernel\_size & padding & layer & in\_features & out\_features  \\
    \hline
    conv2d+relu & 1 & 3 & 3×3 & 1 & linear+swish & 2 & 50 \\
    conv2d+relu & 3 & 6 & 3×3 & 1 & linear+swish & 50 & 200 \\
    conv2d+relu & 6 & 9 & 3×3 & 1 & linear+relu & 200 & 200 \\
    conv2d+relu & 9 & 12 & 3×3 & 1 &  & & \\
    conv2d+relu & 12 & 9 & 3×3 & 1 &  & & \\
    conv2d+relu & 9 & 6 & 3×3 & 1 & & & \\
    conv2d+relu & 6 & 3 & 3×3 & 1 & & & \\
    conv2d+relu & 3 & 1 & 3×3 & 1 & & & \\
    linear & 490 & 200 &  &  & & & \\
    \hline
    \end{tabular}
\end{table*}

\begin{table*}[!htbp]
    \renewcommand{\arraystretch}{1}
    \centering
    \caption{Details of the layers in DeepONet for the inverse problem of Style-A.}
    \label{tab:inverse_layers_style_a}
    \scriptsize
    \begin{tabular}{ccccc|ccc}
    \hline
    \multicolumn{5}{c|}{branch net} & \multicolumn{3}{c}{trunk net} \\
    \hline
    layer & \makecell[c]{in\_channels\\/in\_features} & \makecell[c]{out\_channels\\/out\_features} & kernel\_size & padding & layer & in\_features & out\_features  \\
    \hline
    conv2d+relu & 1 & 5 & 5×5 & 2 & linear+relu & 2 & 200 \\
    conv2d+relu & 5 & 10 & 5×5 & 2 & linear+relu & 200 & 200 \\
    conv2d+relu & 10 & 15 & 5×5 & 2 & linear+relu & 200 & 200 \\
    conv2d+relu & 15 & 20 & 5×5 & 2 &  & & \\
    conv2d+relu & 20 & 15 & 5×5 & 2 &  & & \\
    conv2d+relu & 15 & 10 & 5×5 & 2 &  & & \\
    conv2d+relu & 10 & 5 & 3×3 & 1 & & & \\
    conv2d+relu & 5 & 1 & 3×3 &  & & & \\
    linear+relu & 544 & 200 &  &  & & & \\
    linear & 200 & 200 &  &  & & & \\
    \hline
    \end{tabular}
\end{table*}




\ifCLASSOPTIONcaptionsoff
  \newpage
\fi



%
\bibliographystyle{IEEEtran}

\bibliography{ref}{}

%

\end{document}